
%
%
%
%
\input phyzzx
%
%
\def\eqnon#1{\eqname{#1}}
\def\defeq{\buildrel \rm def \over =}
\def\={\defeq}

\def\sqr#1#2{{\vcenter{\hrule height.#2pt
             \hbox{\vrule width.#2pt height#1pt \kern#1pt
                   \vrule width.#2pt}
             \hrule height.#2pt}}}
\def\square{{\mathchoice{\sqr84}{\sqr84}{\sqr{5.0}3}{\sqr{3.5}3}}}


\def\rmd{{\rm d}}

\def\rmT{{\rm T}}

\def\rmL{{\rm L}}

\def\rms{{\rm s}}

\def\rmsc{{\rm s}}

\def\rmGF{{\rm GF}}

\def\rmFP{{\rm FP}}

\def\wt{\widetilde}

\mathchardef\bfalpha  ="090B
\mathchardef\bfbeta   ="090C
\mathchardef\bfgamma  ="090D
\mathchardef\bfdelta  ="090E
\mathchardef\bfepsilon="090F
\mathchardef\bfzeta   ="0910
\mathchardef\bfeta    ="0911
\mathchardef\bftheta  ="0912
\mathchardef\bfiota   ="0913
\mathchardef\bfkappa  ="0914
\mathchardef\bflambda ="0915
\mathchardef\bfmu     ="0916
\mathchardef\bfnu     ="0917
\mathchardef\bfxi     ="0918
\mathchardef\bfpi     ="0919
\mathchardef\bfrho    ="091A
\mathchardef\bfsigma  ="091B
\mathchardef\bftau    ="091C
\mathchardef\bfupsilon="091D
\mathchardef\bfphi    ="091E
\mathchardef\bfchi    ="091F
\mathchardef\bfpsi    ="0920
\mathchardef\bfomega  ="0921
\REF\GrKo{P.~A.~Griffin and D.~A.~Kosower, Phys.~Lett.~B233~(1989)~295.}
\REF\Va{D.~V.~Vassilevich, Nuovo Cimento A104~(1991)~743.}
\REF\Fu{H.~Fukutaka, Kyoto preprint~YITP/K-897~(1990).}
\REF\FaPo{L.~D.~Faddeev and V.~N.~Popov,
Phys.~Lett.~B25,~29~(1967).}
\REF\FaP{L.~D.~Faddeev, Theor.~Math.~Phys.~1, 1~(1970).}
\REF\Lee{B~.W.~Lee, in {\it Methods in Field Theory}
edited by R.~Balian and J.~Zinn-Justin,
(North Holland, 1975).}
\REF\AbLe{E.~S.~Abers and B~.W.~Lee, Phys.~Rep.~9~(1973)~1.}
\REF\HawZ{S.~W.~Hawking, Commun.~Math.~Phys.~55~(1977)~133.}
\REF\Ber{C.~W.~Bernard, Phys.~Rev.~9~(1974)~3312.}
\REF\HaKu{H.~Hata and T.~Kugo, Phys.~Rev.~D21~(1980)~3333.}
\REF\HaEl{S.~W.~Hawking and G.~F.~R.~Ellis,\
{\it The Large Scale Structure of Space-time}
(Cambridge University Press, Cambridge, 1973).}
\REF\FuP{H.~Fukutaka, in preparation.}
%
\pubnum={YITP/K-958}
\date={January 1992}

\titlepage
\title{Anomalous Jacobian Factor
in the Polyakov~Measure
for~Abelian Gauge~Field~in~Curved~Spacetimes}

\centerline{{\twelverm H}{\tenrm IROKI\ }{\twelverm \ F}{\tenrm UKUTAKA}}
\address{\null\hskip-8mm
Yukawa Institute for Theoretical Physics\break
Kyoto University,~Kyoto 606,~Japan}


\abstract{
The Polyakov measure for the Abelian gauge field is considered
in the Robertson-Walker spacetimes.
The measure is concretely represented
by adopting two kind of decompositions
of the gauge field degrees of freedom
which are most familiarly used
in the covariant and canonical path integrals respectively.
It is shown that the two representations are different
by an anomalous Jacobian factor from each other
and also that the factor has a direct relationship
to an uncancellation factor
of the contributions from the Faddeev-Popov ghost
and the unphysical part of the gauge field
to the covariant one-loop partition function.}
%
%
%
%
%
%
%
\chapter{Introduction}
Euclidean path integrals are most popularly used
as attractive and powerful tools in the investigation of
field theories in curved spacetimes
and quantum gravity.
They enable us to have suggestive discussions with respect to
not only gauge symmetries of systems
if the classical actions have them,
but also manifest coordinate invariance
which matches the spirit of general relativity.

Recently it has been pointed out
in gauge theories in curved spacetimes,
including linearized gravity,
that there is a difference between covariant path integrals
and hamiltonian path integrals, i.e.,
that there is a probability
that covariant path integrals
give rise to uncancellation of one-loop contributions
from the Faddeev-Popov ghosts
and the unphysical degrees of freedom of gauge fields.
In the gravitational case,
the zeta-function calculation of the one-loop partition function
about the $S^4$ saddle point
gives the difference by an integer
in the scaling behavior or $\zeta(0)$ [\GrKo].
In the case of the Abelian gauge field on this manifold
the difference is more complicated [\Va],
and it is also shown
that this difference is not a mere problem of gauge choice,
but it results from a difference
of measure between the covariant and canonical
path integrals [\Fu].

The purpose of this paper is to more clearly show the difference of
the covariant and canonical path-integral measures for vector fields
in the Robertson-Walker spacetimes with $K=+1$.
In order to do it, we study the Polyakov measure and
represent it by using two kind of decompositions
of the gauge field degrees of freedom, i.e.,
the so-called covariant and canonical decompositions,
which may be regarded as suitable ones
for the Lorentz gauge and the Coulomb gauge, respectively
(see Eqs.~(2.3), (2.7) and (3.4)).
Then it may be naively suggested by a formal discussion,
i.e., using the truth
that the Polyakov measure is formally defined
by the Gaussian integral,
that the Jacobian factor under the change of variables
between the two representations would take a trivial value 1.
But in this paper, it is pointed out
that the Jacobian factor may take unfortunately an anomalous value,
which has a direct relationship to the uncancellation factor of
all ghost contributions in the covariant one-loop partition function.

On the other hand, it might be remembered that
connections among path integrals in various gauge choices
could be given by the Faddeev-Popov procedure
[\FaPo-\AbLe].
However, the ordinary discussions are almost too naive.
It is suggested in this paper that
the rigorous discussions in curved spacetimes
must refer to the concrete definition
of the path-integral measure adopted in each of gauge choices,
because the measures may be different from one another.
Therefore it says that all path integrals may not be same.

This paper is organized as follows:
in section 2
the one-loop partition function in the Robertson-Walker spacetimes
is discussed not always
to give the contribution of the only physical modes,
differently from the case of flat spacetime
in which it is well known that
the cancellation of the Faddeev-Popov ghost
and the redundant variables of the gauge field goes well.
In section 3 we concretely construct the Polyakov measure,
using the above mentioned two decompositions,
and then find a nontrivial Jacobian factor
under the transformation between two measures.
Section 4 is devoted to conclusion and discussions,
particularly we discuss with respect to the Faddeev-Popov procedure.
\chapter{One-Loop Partition Function in Robertson-Walker Spacetimes}

In order to discuss
that the covariant path integral in curved spacetimes has
a probability that
the one-loop contribution of the Faddeev-Popov ghost
does not cancel out with one
from the unphysical gauge field,
we study, in this section, the case of the Robertson-Walker spacetimes
with actually setting up a special coordinate system.

First, let us start with the BRST Euclidean path integral
in any $D$-dimensional curved spacetime;
$$
  {\cal Z}
 \=
  \int{\cal D}A_{I}\Delta_{\rmFP}
  \exp\Big[ -{1\over\hbar}(I+I^\rmGF) \Big]
\eqn\OLC
$$
with
$$\eqalign{
  I
&\=
  \int \rmd^Dx
  {1\over 4}\sqrt gg^{IM}g^{JN}F_{IJ}F_{MN},
\cr
  I^\rmGF
&\=
  \int \rmd^Dx
  {1\over2\alpha}\sqrt g(\nabla^IA_I)^2,
\cr
  \Delta_{\rmFP}
&\=
  \Big\vert{1\over\sqrt\alpha}\det-\square(s)\Big\vert,
\cr
}\eqn\WOLC$$
where $s$ means that the covariant d'Alembertian $\square$
acts on scalar fields
and the functional measure is defined
by the same method as that used in Ref.~\HawZ\
so as to have coordinate invariance.
Now the gauge field $A_I$ can be decomposed into
a divergenceless vector $A^\rmd_I$
and a covariant derivative of a scalar field, $A^\rms_I$;
$$
  A_I
 =
  A_I^{\rmd}+A_I^{\rmsc},
\qquad
  \nabla^IA_I^{\rmd}
 =
  0,
\qquad
  A_I^{\rmsc}
 \=
  \nabla_I\square^{-1}S,
\eqn\DecAI
$$
and then these two parts are completely decoupled
in $I$ and $I^\rmGF$,
because the gauge invariant action $I$ is independent of $A^\rms_I$
since $S$ can be considered a parameter
of the gauge transformation and
$I^\rmGF$ is obviously independent of $A^\rmd_I$.
Thus the one-loop determinant of $A_I$
in \OLC\ is made of completely separated contributions
of these two parts, i.e.,
$
  {\cal Z}^{(\rmd)}
  \times
  {\cal Z}^{(\rmsc)}
$
in which ${\cal Z}^{(\rmd)}$ and ${\cal Z}^{(\rmsc)}$ are
one-loop corrections from $A^\rmd_I$ and $A^\rms_I$ respectively;
$$\eqalign{
  {\cal Z}^{(\rmd)}
&=
  \Big\vert\det\Big[ -\square g^{IJ} +R^{IJ}
  \Big](v^\rmd)\Big\vert^{-1/2},
\qquad
  {\cal Z}^{(\rmsc)}
 =
  \Big\vert\det-\alpha^{-1}\square(s)\Big\vert^{-1/2},
\cr
}\eqn\OLCAISc
$$
where $v^\rmd$ denotes divergenceless vector fields.
Therefore, putting them in \OLC,
${\cal Z}$ is written in a simple form
usable in any curved spacetime:
$$
  {\cal Z}
 =
  \Big\vert\det-\square(s)\Big\vert^{1/2}
  \times{\cal Z}^{(\rmd)}.
\eqn\FormOLC
$$
It is important to note here
that the contribution of the gauge parameter,
${\cal Z}^{(\rms)}$, always cancels out
with one from a half degree of freedom
of the Faddeev-Popov ghost field
(of course, we must take the same boundary condition
on the fermionic Faddeev-Popov ghost as
that on the bosonic gauge field,
similarly to the case of finite temperature gauge theories
[\Ber,\HaKu]),
however, it is not obvious
whether the contribution of the longitudinal part,
which is a gauge invariant but unphysical part
of the divergenceless vector, cancels out
with the remaining contribution of the Faddeev-Popov ghost field
and also whether the one-loop partition function \OLC\
gives the contribution
from the only physical modes, i.e., the transverse part
of the divergenceless vector.

Next, in order to study this issue in detail,
let us study the case of
the Robertson-Walker spacetimes with $K=+1$,
choosing coordinates
so that the metric has the form
$$
  \rmd s^2
 =
  \rmd \tau^2 +a^2(\tau)\rmd\Omega_{D-1}^{\ 2},
\qquad
  \rmd\Omega_{D-1}^{\ 2}
 \=
  {\wt g}_{ij}\rmd x^i\rmd x^j,
\eqn\LERW
$$
where $\wt g_{ij}$ is
the metric of the unit $(D-1)$-sphere $S^{D-1}$
according to $K=+1$ [\HaEl].
In such coordinates, regarding $\tau$ as a time coordinate,
$A_I^\rmd$ can be split
into transverse and longitudinal components;
$$\eqalign{
&\hskip 2.5cm
  A_I^\rmd
 =
  A_I^\rmT+A_I^\rmL,
\cr
&
  A_D^\rmT
 =
  \wt\nabla^iA_i^\rmT
 =
  0,
\qquad
  A_i^\rmL
 \=
  -\wt\nabla_i\wt\square^{-1}
  a^{-(D-3)}\partial_D(a^{D-1}A_D^\rmL),
\cr
}\eqn\DefAId
$$
where $\wt\nabla_i$ and $\wt\square$ are
the covariant derivative and the covariant Laplacian
on the unit $S^{D-1}$.
We then find the separation of two parts in $I^\rmd$,
which means that
${\cal Z}^{(\rmd)}$ becomes
${\cal Z}^{(\rmT)}\times{\cal Z}^{(\rmL)}$ with
$$\eqalign{
  {\cal Z}^{(\rmT)}
&=
  \Big\vert\det\Big[ -\square g^{IJ} +R^{IJ}
  \Big](v^\rmT)\Big\vert^{-1/2},
\cr
  {\cal Z}^{(\rmL)}
&=
  \Big\vert\det\Big[ -\square
  +{D-2\over D-1}R_{DD}
  \Big](\wt s)\Big\vert^{-1/2},
\cr
}\eqn\OLCAITra
$$
where $v^\rmT$ and $\wt s$, respectively, are transverse vector fields
and scalar fields having no zero-eigenvalue mode of $\wt\square$,
and ${\cal Z}^{(\rmL)}$ is easily obtained
by the aid of the following formulae
satisfying under certain boundary conditions;
$$\eqalignno{
  \int\rmd^Dx
  \sqrt g g^{IJ}A_I^\rmL A'^\rmL_J
&=
  \int\rmd^Dx
  \sqrt g
  (aA_D^\rmL)F^{-2}(aA'^\rmL_D),
&\eqnon\VIWIRW\cr
  \int\rmd^Dx
  \sqrt gA_I^{\rmL}
  \Big[ -\square g^{IJ} +R^{IJ} \Big]A'^\rmL_J
&=
  \int\rmd^Dx
  \sqrt g
  (aA_D^\rmL)(-\wt\square)F^{-2}F^{-2}(aA'^\rmL_D),
\cr
}$$
where $F$ is defined by
$$\eqalignno{
  F
&\=
  \Big(
  (-\wt\square)
  \Big[
  -\square^{(\rms)}+{D-2\over D-1}R_{DD}
  \Big]^{-1}
  \Big)^{1/2}
&\eqnon\DefF\cr
}$$
with
$
  R_{DD}
 =
  -(D-1)a^{-1}\ddot a
$
and the d'Alembertian operator acting on
scalar fields, $\square^{(\rms)}$.
It must be noted here that $F$ is not commutable
with $a$ and $\partial_D$. From Eqs.~\FormOLC\ and \OLCAITra,
therefore, we obtain
$$\eqalignno{
  {\cal Z}
&=
  \Big\vert\det-\square(s)\Big\vert^{1/2}
  \Big\vert\det\Big[ -\square +{D-2\over D-1}R_{DD}
  \Big](\wt s)\Big\vert^{-1/2}
  {\cal Z}^{(\rmT)}.
&\eqnon\GOLCRW\cr
}$$

As the conclusion in this section,
the extra factor in Eq.~\GOLCRW, i.e.,
$$\eqalignno{
&
  \Big\vert\det-\square(s)\Big\vert^{1/2}
  \Big\vert\det\Big[ -\square +{D-2\over D-1}R_{DD}
  \Big](\wt s)\Big\vert^{-1/2},
&\eqnon\HSF\cr
}$$
means that
${\cal Z}$ in Robertson-Walker spacetimes
is not the one-loop contribution
which the only physical modes give
if it does not become a trivial value 1
under a used regularization,
so the factor \HSF\ becomes the uncancellation factor
which expresses the contributions from the Faddeev-Popov ghost and
the unphysical gauge field.
Indeed, we know that in de~Sitter spacetime it happens,
i.e., the zeta-function calculation does not make it 1 $[\Va]$.

\chapter{Anomalous Jacobian Factor in
the Polyakov Measure}

In this section we concretely construct
the Polyakov measure for vector fields
in the Robertson-Walker spacetimes, which is formally defined by
$$\eqalignno{
  \int{\cal D}A_I
  \exp\Big[ -{1\over2\hbar}< A,A > \Big]
&\=
  1
&\eqnon\DefPM\cr
}$$
with
$$\eqalignno{
  < A,A' >
&\=
  \int\rmd^Dx\sqrt g
  g^{IJ}A_IA'_J.
&\eqnon\DefSIP\cr
}$$
The decompositions used in the discussion here are the following two:
the first decomposition, which is covariant one, is
Eqs.~\DecAI\ and \DefAId\ with
$$\eqalign{
  A^{\rmL}_D
&\=
  a^{-1}FY,
}\eqn\DefALi
$$
where $Y$ is introduced as a scalar field
in the Robertson-Walker spacetimes,
and the second decomposition, being so-called canonical one, is
represented by
$$\eqalign{
  A_I
&=
  A^{A_D}_I+A^\rho_I+A^{\rmT}_I,
\cr
  A^{A_D}_D
&\=
  A_D,
\qquad
  A^{A_D}_i
 \=
  0,
\cr
  A^{\rho}_D
&\=
  0,
\quad\qquad
  A^{\rho}_i
 \=
  \wt\nabla_i\wt\square^{-1}\rho.
\cr
}\eqn\FDec
$$

The Gaussian integral is
$
  \int_{-\infty}^{+\infty}
  \rmd x e^{-\lambda x^2}
 =
  \sqrt{\pi/\lambda},
$
thus using the covariant decomposition leads to
the covariant measure;
$$\eqalignno{
&
  \int
  {{\cal D} S\over\sqrt{2\pi\hbar}}
  {{\cal D} Y\over\sqrt{2\pi\hbar}}
  {{\cal D} A^\rmT_I\over\sqrt{2\pi\hbar}}
  \exp\Big[ -{1\over2\hbar}< A,A > \Big]
 =
  1,
&\eqnon\NINPM\cr
}$$
where ${\cal D}S$, ${\cal D}Y$ and ${\cal D}A_I^\rmT$ may be defined
with expansion coefficients of them
in an orthonormal and complete set in the $D$-dimensional spacetime,
and if we use the canonical decomposition
the Polyakov measure is defined by
the canonical one $[\Fu]$;
$$\eqalign{
&
  \int\prod_{\tau}
  \Big[
  \Big( {\Delta\tau a^{D-1}(\tau)\over2\pi\hbar} \Big)^{1/2}
  {\cal D}A_D(\tau)
  \Big]
  \prod_{\tau}
  \Big[
  \Big(
  {\Delta\tau a^{D-3}(\tau)\over2\pi\hbar\big(-\wt\square(\wt s)\big)}
  \Big)^{1/2}
  {\cal D}\rho(\tau)
  \Big]
\cr
&\quad
  \times
  \prod_{\tau}
  \Big[
  \Big( {\Delta\tau a^{D-3}(\tau)\over2\pi\hbar} \Big)^{1/2}
  {\cal D}A^{\rmT}_i(\tau)
  \Big]
  \exp\Big[ -{1\over2\hbar}< A,A > \Big]
 =
  1,
\cr
}\eqn\InPM
$$
in which the time $\tau$ is specialized from the other coordinates,
and its product might be defined in the discrete time formulation
with a finite distance
and its zero limitation after integrations,
and then the functional measures on each time,
${\cal D}A_D(\tau)$, ${\cal D}\rho(\tau)$
and ${\cal D}A_i^\rmT(\tau)$, are defined
with their expansion coefficients
in the basis of eigenfunctions of $\wt\square$
on the unit $S^{D-1}$.
Furthermore, it is noted here that
$
  \rho
 =
  {\wt \nabla}^iA_i
$
has no zero mode owing to
$$
  \int\rmd^{D-1}x\sqrt{\wt g}
  {\wt \nabla}^iA_i{\wt S}^{0m}
 =
  0,
\eqn\ScrhoZM
$$
where ${\wt S}^{0m}$ is an eigenfunctions
with zero eigenvalue of $\wt\square$.

As mentioned in section 1, Eqs.~\NINPM\ and \InPM\ naively mean that
the Jacobian factor under the change of variables
between their two measures
would become trivially 1,
since both measures are defined through the Gaussian integral
whose integration value is 1.
However, as discussed below, the factor is not 1
and take an anomalous value.

First, we study about the physical variable:
its relation between the two decompositions
tells us that
$$\eqalign{
  \prod_\tau
  \Big[
  \Big( {\Delta\tau a^{D-3}(\tau)\over2\pi\hbar} \Big)^{1/2}
  {\cal D}A^{\rmT}_i(\tau)
  \Big]
&=
  {{\cal D} A^\rmT_I\over\sqrt{2\pi\hbar}},
\cr
}\eqn\DefMAT
$$
where we note that the time $\tau$ product in the l.h.s.~is changed,
in the r.h.s.,
into the product of modes along the time axis,
because we may more expand
$A^\rmT_i(\tau)$, which is already expanded
with eigenfunctions on $S^{D-1}$,
by using eigenfunctions
in the Robertson-Walker spacetimes.
Furthermore,
when expressing eigenfunctions in the Robertson-Walker Spacetimes
in terms of eigenfunctions on $S^{D-1}$
and certain functions of the time $\tau$,
the factors,
$\prod_\tau\big( \Delta\tau a^{D-3}(\tau) \big)^{1/2}$,
cancel out with determinants of the time functions
(see Ref.~\Fu~in the $S^D$ case).
Eq.~\DefMAT, therefore, means that, with respect to the physical part,
no anomalous thing happens
under the transformation between the covariant and canonical measures.

Next, as for the unphysical variables,
their relations between the two decompositions become
$$\eqalignno{
  A_D
&=
  \partial_D(-\square)^{-1/2}S
  +a^{-1}FY,
&\eqnon\KanAD\cr
  \rho
&=
  \wt\square(-\square)^{-1/2}S
  -a^{-(D-3)}\partial_D( a^{D-2}FY ),
&\eqnon\Kanrho\cr
  S
&=
  -(-\square)^{-1/2}a^{-(D-1)}\partial_D( a^{D-1}A_D )
  -(-\square)^{-1/2}a^{-2}\rho,
&\eqnon\KanS\cr
  Y
&=
  Fa^{-1}A_D
  -Fa^{-1}\wt\square^{-1}\partial_D\rho,
&\eqnon\KanY\cr
}$$
where the relation between the former two equations and
the later two equations is that of the inverse transformation.
Now, let us separate the unphysical variables
into the zero-mode and nonzero-mode parts of $\wt\square$,
and study the Jacobian factor in each part.

With respect to zero mode of $\wt\square$,
using Eq.~\KanAD\
gives
$$\eqalignno{
  \prod_\tau
  \Big[
  \Big( {\Delta\tau a^{D-1}(\tau)\over2\pi\hbar} \Big)^{1/2}
  {\cal D}A_D^{(\wt0)}(\tau)
  \Big]
&=
  \vert\det[{\partial_D}{(\wt0)}]\vert
  \Big\vert\det-\square{(\wt0)}\Big\vert^{-1/2}
  {{\cal D} S^{(\wt0)}\over\sqrt{2\pi\hbar}},
&\eqnon\KanPIZ\cr
}$$
where the symbol $(\wt0)$ means
that the values are of the zero-mode part,
and we used the truth that $\rho$ has no zero mode
and the assumption that $Y$, too, does not have the mode
to be consistent
with the $A^\rmL_I$ definition. From Eq.~\KanPIZ, thus, if the factor,
$$
  \vert\det[{\partial_D}^{(\wt0)}]\vert
  \Big\vert\det-\square^{(\wt0)}\Big\vert^{-1/2},
\eqn\FacZ
$$
is 1, which is of course satisfied in flat spacetime,
the two measures, i.e.,
the covariant and canonical measures,
are same in this zero-mode part of $\wt\square$.

As for the nonzero-mode part of $\wt\square$,
we may find a nontrivial Jacobian factor
having a direct relationship to the uncancellation factor
\HSF.
In order to derive it,
let us carry out two kind of calculations:
first, using \KanAD\ and \KanS\ at the following steps;
$
  (A_D,\rho)
 \rightarrow
  (A_D,S)
 \rightarrow
  (Y,S)
$,
then we have
$$\eqalignno{
&\quad
  \prod_\tau
  \Big[
  \Big( {\Delta\tau a^{D-1}(\tau)\over2\pi\hbar} \Big)^{1/2}
  {\cal D}A_D(\tau)
  \Big]
  \Big[
  \Big(
  {\Delta\tau a^{D-3}(\tau)\over2\pi\hbar
  \big( -\wt\square(\wt s) \big)}
  \Big)^{1/2}
  {\cal D}\rho(\tau)
  \Big]
\cr
&=
  \Big\vert\det-\square(\wt s)\Big\vert^{1/2}
  \Big\vert\det
  \Big[-\square+{D-2\over D-1}R_{DD}\Big](\wt s)
  \Big\vert^{-1/2}
  {{\cal D} S\over\sqrt{2\pi\hbar}}
  {{\cal D} Y\over\sqrt{2\pi\hbar}},
&\eqnon\AJa\cr
}$$
while if we use Eqs.~\Kanrho\ and \KanY\ at the steps such that
$
  (A_D,\rho)
 \rightarrow
  (Y,\rho)
 \rightarrow
  (Y,S)
$,
we obtain
$$\eqalignno{
&\quad
  \prod_\tau
  \Big[
  \Big( {\Delta\tau a^{D-1}(\tau)\over2\pi\hbar} \Big)^{1/2}
  {\cal D}A_D(\tau)
  \Big]
  \Big[
  \Big(
  {\Delta\tau a^{D-3}(\tau)\over2\pi\hbar
  \big( -\wt\square(\wt s) \big)}
  \Big)^{1/2}
  {\cal D}\rho(\tau)
  \Big]
\cr
&=
  \Big\vert\det-\square(\wt s)\Big\vert^{-1/2}
  \Big\vert\det
  \Big[-\square+{D-2\over D-1}R_{DD} \Big](\wt s)
  \Big\vert^{1/2}
  {{\cal D} S\over\sqrt{2\pi\hbar}}
  {{\cal D} Y\over\sqrt{2\pi\hbar}}.
&\eqnon\AJb\cr
}$$
{}From Eqs.~\AJa\ and \AJb, therefore, the factor,
$$
  \Big\vert\det-\square(\wt s)\Big\vert^{1/2}
  \Big\vert\det
  \Big[-\square+{D-2\over D-1}R_{DD} \Big](\wt s)
  \Big\vert^{-1/2},
\eqn\FacA
$$
might be thought to become naively 1 and both measures would be
concluded to equal each other
also as to the non-zero part of $\wt\square$.
But the factor is obviously the same with the uncancellation factor
\HSF\ in the covariant one-loop partition function
excluding the zero-mode part,
and in order to actually calculate the factor \FacA\
we need a regularization
because the arguments of the determinants
are infinite matrices.
Thus, some regularizations might fail to make its value 1.
Indeed, in the case of de~Sitter spacetime,
the zeta-regularization calculation does not make it 1,
as mentioned in section 2.
Hence the factor \FacA\ may be called an anomalous Jacobian factor
under the change of variables
between the covariant and canonical measure.

\chapter{Conclusion and Discussions}
In this paper it was shown that in the Robertson-Walker spacetimes
the path integral formula defined with the covariant measure
does not provide the one-loop partition function
which is made of contributions of the only physical modes
and there the cancellation of the one-loop contributions
between the Faddeev-Popov ghost and
the unphysical modes of the gauge field does not hold,
similarly to the case of de~Sitter spacetime.
We also studied about the Polyakov measure
by representing it in terms of two kind of decompositions,
i.e., the covariant and canonical ones,
and showed that an anomalous Jacobian factor,
which might not be 1 in some cases of regularizations,
exists under the transformation between the two measures
and it is the same with the above uncancellation factor
in the one-loop partition function.

Let us discuss the regularization dependence
in the anomalous Jacobian factor.
If we take the discrete time formulation throughout,
the Jacobian factor might become 1.
Since the number of variables along the time axis is finite
in the formulation,
although degrees of freedom with respect to the spatial part
are infinite,
so the change of variables between the two Gaussian integrals,
\NINPM\ and \InPM, is well-defined.
Hence the naive discussion can be adopted without any problem.
Of course,
the actual calculation of the Jacobian factor
is very difficult in curved spacetimes,
though
the arguments of the determinants become finite matrices.
This discussion may suggest that
there is a regularization dependence in the Jacobian factor,
in particular, we may note that
the one-loop calculation with the discrete time formulation
might have a difference from
the zeta-function calculation.

It is known that,
also
in the linearized gravitational case about de~Sitter spacetime,
the complete cancellation of the one-loop contributions
from the all redundant degrees of freedom
does not go well. Furthermore, the uncancellation factor
may be related with an anomalous Jacobian factor
like the case of the Abelian gauge field $[\FuP]$.

Finally we turn our discussion to the Faddeev-Popov procedure.
It might be believed that
their procedure can make all path-integral formulae
defined in various gauge choices
be connected.
However, as being obvious from section 3,
how to define the path-integral measure used in each gauge fixing
is important to have a rigorous discussion
with respect to the dependence of gauge choice in path integrals
in curved spacetimes.
Because in order to make delta-functions
used in the Faddeev-Popov procedure be well-defined,
we have to adopt
the most suitable decomposition as their arguments.
For example, in the case of the Coulomb gauge
the canonical decomposition is best,
on the other hand, the covariant decomposition is most suitable
for the Lorentz gauge case.
Therefore it may be concluded that
we must pay attention to that what kind of the decomposition
of the gauge field degrees of freedom is used
to define the path-integral measure
whenever the Faddeev-Popov procedure is applied to a rigorous discussion
of the gauge dependence in path integrals.

\refout
\bye